\documentclass[twocolumn,
showpacs,aps,prb,amsmath,amssymb]{revtex4}
\usepackage{graphicx}

\begin{document}

\title{Verwey transition in Fe$_{3}$O$_{4}$ at high pressure: quantum
critical behavior at the onset of metallization}

\author{J.~Spa\l{}ek$^{1,2}$}
\email{ufspalek@if.uj.edu.pl}
\author{A.~Koz\l{}owski$^{2}$}
\email{kozlow@agh.edu.pl}
\author{Z.~Tarnawski$^{2}$ }
\author{Z.~K\c{a}kol$^{2}$}
\author{Y.~Fukami$^{3}$}
\author{F.~Ono$^{3}$}
\author{R. Zach$^{4}$}
\author{L.~J.~Spalek$^{5}$}

\affiliation{$^{1}$Marian Smoluchowski Institute of Physics, Jagiellonian
University, Reymonta 4, 30-059 Krak\'ow, Poland}
\affiliation{$^{2}$Department of Solid State Physics, AGH University
of Science and Technology,
Al. Mickiewicza 30, 30-059 Krak\'ow, Poland}
\affiliation{$^{3}$Department of Physics, Okayama University,
Okayama 700-8530, Japan}
\affiliation{$^{4}$ Institute of Physics, University
of Technology, Podchor\c{a}\.zych 1, 30-084 Krak\'ow, Poland, }
\affiliation{$^{5}$Quantum Matter Group, Cavendish Laboratory,
University of Cambridge, Cambridge CB3 0HE, UK}
\date{\today}
\begin{abstract}
We provide evidence for the existence of a {\em quantum critical
point\/} at the metallization of magnetite Fe$_{3}$O$_{4}$ at an
applied pressure of $p_{c} \approx 8$~GPa. We show that the present
ac magnetic susceptibility data support earlier resistivity data.
The Verwey temperature scales with pressure $T_{V}\sim
(1-p/p_{c})^{\nu}$, with $\nu\sim 1/3$. The resistivity data shows a
temperature dependence $\rho(T)=\rho_{0}+AT^{n}$, with $n\simeq 3$
above and $2.5$ at the critical pressure, respectively. This
difference in $n$ with pressure is a sign of critical behavior at
$p_{c}$. The magnetic susceptibility is smooth near the critical
pressure, both at the Verwey transition and near the ferroelectric
anomaly. A comparison with the critical behavior observed in the
Mott-Hubbard and related systems is made.
\end{abstract}
\maketitle


Magnetite (Fe$_{3}$O$_{4}$) is among the oldest and most fascinating
materials throughout the history of science. It is ferrimagnetic
below the Curie temperature $T_{C}\approx 860$~K and semiconducting
with a room temperature resistivity of $\rho(T=300$~K$)\simeq
10^{-2}\Omega$cm. The stoichiometric material undergoes a
discontinuous Verwey transition\cite{1} to a higher-resistivity
phase at a temperature of $T_{V}\simeq (121\pm 1)$~K, at which the
resistivity at ambient pressure jumps up by about two orders of
magnitude upon cooling. The transition is of a
semiconductor-semiconductor type at lower pressures,
although the exact difference between them remains somewhat elusive.
In addition, the net magnetic moment has the value $\mu\simeq
4.1\mu_{B}$ per formula and approximately corresponds to the spin
$S=2$ in Fe$^{2+}$ high-spin configuration in the octahedral
position (the remaining two Fe$^{3+}$ ions, one in tetrahedral and
one in the octahedral positions with spin $S=5/2$ have their moments
antialigned). The high-spin state is induced by Hund's rule coupling
among $3d$ electrons.

Essential progress was made when samples with an extremely well
controlled composition were synthesized\cite{2} and studied.\cite{3}
Notably, the change of the Verwey transition from first- to
second-order has been discovered, depending on the off-stoichiometry
parameter $\delta$ of Fe$_{3-\delta}$O$_{4}$. Also, the
metallization of Fe$_{3}$O$_{4}$ has been achieved\cite{4} at a
pressure of $p_{c}\simeq 8$~GPa.

The primary purpose of this work is to systematize the change in the
properties of Fe$_{3}$O$_{4}$ with pressure and to single out
universal features of these results. During the course of this work
we reanalyzed the earlier resistivity data\cite{4} and have
discovered that the combined magnetic susceptibility and resistivity
data point to the possibility of a {\em quantum critical point\/}
existing at the metallization threshold. This is explicitly verified
by determining the pressure dependence of the critical exponents of
both the transition temperature $T_{V}(p)$ and the resistivity just
above the transition $\rho(T_{V+})$, as well as that of $\rho(T)$ at
the critical pressure and above it. Also, the pressure dependence of
the two low-temperature anomalies in the ac susceptibility,
presumably connected with the appearance of ferroelectricity, evolve
with increasing pressure, but this aspect of the work will not be
detailed here.

The importance of the presence of a quantum-critical-point (QCP) at
the localization- delocalization boundary cannot be underestimated
as there are no well defined {\em universality classes\/} as yet for
this electronic transition. Previously, a {\em classical\/} critical
point (of the Van der Waals-type) was determined\cite{5} in a
three-dimensional system Cr-doped V$_{2}$O$_{3}$, at
$T_{crit}=457.5$~K and $p_{crit}\simeq 3.7$~kbar, the existence of
which was suggested earlier.\cite{6} The critical behavior in this
case has the critical exponents of the mean-field type, as the
$T_{crit}$ value is rather high. Recently, a critical point has been
seen\cite{7} in a quasi-two-dimensional organic solid at
$T_{crit}=39.7$~K and $p_{c}=24.8$~MPa, with completely different
critical exponents than those for the above three-dimensional case.
A true quantum critical point (at $T=0$) at the
localization-delocalization boundary has been detected at the Kondo
semiconductor - non-Fermi liquid boundary upon chemical substitution
in CeRhSb$_{1- x}$Sn$_{x}$.\cite{8} In the last system, it is
connected with the delocalization of $4f$ electrons due to Ce$^{3+}$
and is induced by the change in carrier concentration upon doping.
We suggest, that the present system (Fe$_{3}$O$_{4}$) represents the
first example of a system with a true {\em quantum critical point\/}
(with $T_{crit}=0$) in a monocrystalline and stoichiometric system
in a spin-polarized state, both in the semiconducting and the
metallic states. In contrast, a different type of behavior has been
observed\cite{9,10} in the antiferromagnetic system
NiS$_{2-x}$Se$_{x}$, in which the Hund's rule coupling, leading to
the high-spin $S=1$ state in this case, also occurs. The differences
are caused by a disparate magnetic ordering, as well as by the
circumstance that, as we show, the metallic state of Fe$_{3}$O$_{4}$
is that of a {\em non-Fermi liquid\/}, whereas that of
NiS$_{2-x}$Se$_{x}$ is an almost localized antiferromagnetic Fermi
liquid.

A single crystalline magnetic sample used for the ac magnetic
susceptibility measurements, was cut from a larger crystal grown
from the melt using the cold crucible technique (skull
melter)\cite{2}, at Purdue University. The crystal was then annealed
under CO/CO$_{2}$ gas mixtures to establish the
stoichiometry\cite{2} and rapidly quenched to room temperature to
freeze in the high temperature thermodynamic equilibrium. Although
this procedure generates octahedral defects, most of the low
temperature electronic processes remain intact, as is evidenced by
the sharp Verwey transition temperature $T_{V}$ and the existence of
the low tempearature anomaly (observed only in the best crystals).

\begin{figure}
\includegraphics[width=0.4\textwidth]{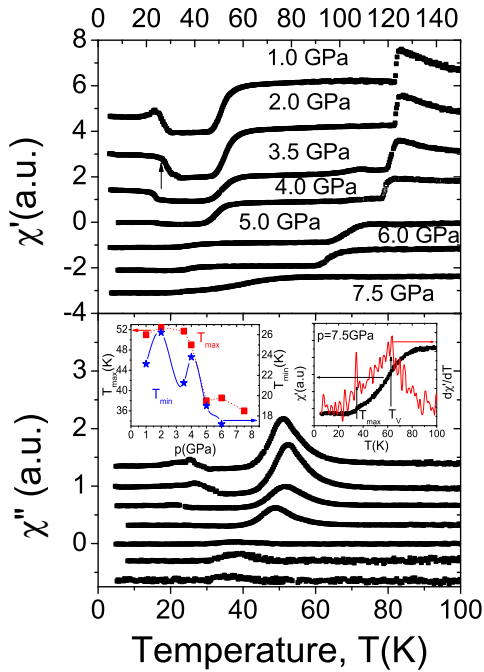}
\caption{\label{fig:f1} Pressure dependence of ac magnetic
susceptibility (at a frequency of $120$Hz) for the stoichiometric
single crystal; note different $T$ scales for the real
($\chi^{\prime}$) and the imaginary ($\chi^{\prime\prime}$) parts of
$\chi$. Right inset: illustration of the definition of $T_{V}$ from
maximum of $d\chi^{\prime}/dT$ when the sharp step in
$\chi^{\prime}$ was not observed (i.e. for $p>3.5$~GPa). Left inset:
$T_{max}$ indicates the position of the anomaly near $50$~K, whereas
$T_{min}$ the lower anomaly (indicated by the vertical arrow).}
\end{figure}

In Fig. 1 we display the temperature dependence of the ac
susceptibility $\chi=\chi^{\prime}+i\chi^{\prime\prime}$ components
for selected pressures. The upper panel provides the real part
$\chi^{\prime}$, for which three characteristic temperatures are:
The Verwey temperature $T_{V}$ (the uppermost jump), and the {\em
two temperatures associated with ferroelectricity\/},\cite{11} for
$T_{max}\sim 60-50$~K, and the lower anomaly at $T_{min}\sim 20$~K.
With increasing pressure, all of them shift towards lower
temperatures and eventually disappear for pressures $p\gtrsim
7.5$~GPa. Note that pronounced losses are observed near $T_{max}$.
They are most probably due to the ferroelectric transition and
associated with it ferroelectric domain dynamics. In the insets we mark
the temperatures $T_{V}$, $T_{max}$, and $T_{min}$ near the critical
pressure. The first two temperatures approach each other, as well as
decrease with increasing pressure.

\begin{figure}
\includegraphics[width=0.4\textwidth]{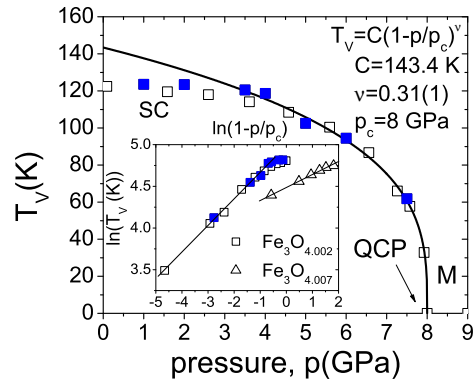}
\caption{\label{fig:f2} Pressure dependence of the Verwey transition temperature $T_{V}$. We
mark the data points from this work as solid squares (blue online) and those
of M$\hat{\mbox{o}}$ri et al.\cite{4} by open squares. Solid line represents the
fit specified and ending at quantum critical point (QCP). Inset:
same in the doubly
logarithmic scale. Open triangles show the respective critical
behavior for slightly nonstochiometric ($\delta=0.007$) sample. M denotes
a stable metallic state,
whereas SC labels the semiconducting state below $T_{V}$.}
\end{figure}

To explicitly demonstrate the evolution of the Verwey transition
with pressure, we plot (Fig. 2) the $T_{V}(p)$ dependence by
combining our magnetic susceptibility results and the resistivity
data taken from Ref. \onlinecite{4}. In the inset we redraw the data
on a doubly logarithmic scale for both the almost perfectly
stoichiometric (Fe$_{3}$O$_{4.002}$) and the slightly
nonstoichiometric (Fe$_{3}$O$_{4.007}$) samples to demonstrate the
exponential fit quality, as well as to show the difference between
the two cases. We observe a clear exponential behavior of
$T_{V}(p)=A(1-p/p_{c})^{\nu}$ above $3.5$~GPa for the stoichiometric
sample, with the critical pressure $p_{c}\simeq 8$~GPa and the value
of the exponent $\nu\simeq 1/3$. Actually, the $\rho(T)$ data of
M$\hat{\mbox{o}}$ri et al.\cite{4} already shows metallic
conductivity at $p=7.9$~GPa down to about $2$~K, but a percent
inaccuracy in determining $p_{c}$ is acceptable given the inaccuracy
of determining $T_{V}$ for $p\gtrsim 7$~GPa. As the data in the
inset shows, both $p_{c}$ and $\nu$ depend strongly on the sample
stoichiometry. For example, for $\delta =0.007$ we have $p_{c}\simeq
6.1$~GPa and the critical exponent is reduced to the value
$\nu\simeq 1/6$. This means that the nonstoichiometry {\em is not\/}
equivalent to a reduction of critical pressure.

\begin{figure}
\includegraphics[width=0.4\textwidth]{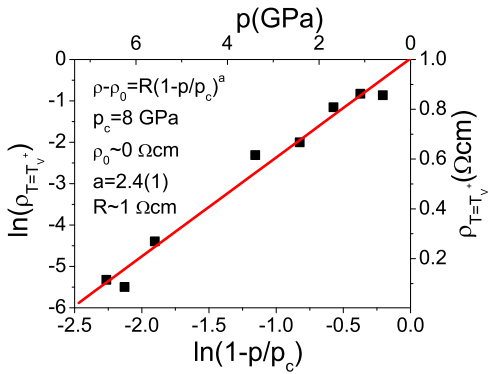}
\caption{\label{fig:f3} Pressure dependence of the electrical resistivity just above the Verwey
transition\cite{4} (at $T=T_{V+}$). The residual resistivity obtained from the
fit is
$\rho_{0}\simeq 0$ within the fitting error.}
\end{figure}

The nature of the metallic
state $(M)$ above $p_{c}$ should be characterized in greater detail. This is
particularly important because for $p< 7.5$~GPa the state
{\em above\/} $T_{V}$ is clearly a semiconductor.
A similar type of behavior to that exhibited in Fig. 2,
though without any reference to
{\em quantum critical behavior\/}, was suggested
by Rosenberg et al.\cite{12}
Also, the magnetic susceptibility and the resistivity
discontinuities in Fig. 2 follow
the same dependence $T_{V}(p)$.

\begin{figure}
\includegraphics[width=0.4\textwidth]{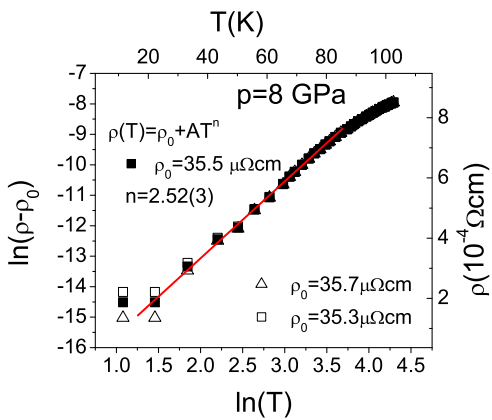}
\caption{\label{fig:f4} Temperature dependence of electrical resistivity
$\rho(T)=\rho_{0}+AT^{n}$
of Fe$_{3}$O$_{4.002}$ at $p_{c}=8$~GPa. The
solid line ({\em red online\/}) corresponds to $\rho_{0}=35.5\,\mu\Omega$cm and
$n=2.52$.
Open symbols show the sensitivity of the fits to $\rho_{0}$ value:
slight variations of the $\rho_{0}$ value do not appreciably
change the fitting accuracy.}
\end{figure}

To provide quantitative evidence that we observe a criticality near
the $T_{crit}=0$, $p=p_{c}$ point, one requires a proper definition
of the order parameter, as well as its dependence on $p-p_{c}$. In
the previous studies of criticality near the insulator-metal
transition\cite{5,7} the role of the order parameter was played by
the conductivity difference $\sigma(p) - \sigma(p_{c})$ or
equivalently, directly by the resistivity.\cite{9} The two transport
quantities mimic the essential difference between carrier
concentrations on both sides of the electronic transition in the
equilibrium state. Additional evidence is provided by the behavior
of physical quantities near $p_{c}$. For that purpose, our data from
Fig. 1 is not directly useful, as the susceptibility is continuous
upon approaching the system metallization. This in turn means, that
magnetism is robust under the applied pressure, as no clear
signature of {\em critical magnetic behavior\/} has been observed up
to $300$~K. In view of that circumstance, we reanalyzed the data of
M$\hat{\mbox{o}}$ri et al.\cite{4} and plot in Figs. 3 and 4,
respectively, the resistivity $\rho(T_{V+})$ just above the Verwey
transition as a function of $p-p_{c}$, as well as $\rho(T)$ at
$p_{c}$. We observe that, in the former situation we have roughly
the scaling $\rho(p)=R[( p_{c}-p)/p_{c}]^{a}$, with $R\simeq
1\,\Omega$cm and $a\simeq 5/2$, whereas at $p=p_{c}$:
$\rho(T)=\rho_{0}+AT^{n}$, where $\rho_{0}\simeq 35.5\mu\Omega$cm,
and $n\simeq 5/2$. These results differ from those for Mott-Hubbard
systems\cite{5,7,9} obtained when the critical temperature
$T_{crit}>0$. However, it should be underlined again that, as only
in the present case $T_{crit}=0$, we are dealing with a true quantum
critical regime for $p\rightarrow p_{c}$.

In Fig. 5 we plot the temperature dependence of the resistivity in
the metallic phase (for $p=9$~GPa). The resistivity $\rho(T)$ is
roughly proportional to $T^{3}$ and the value of the residual
resistivity is considerably reduced. The reduction of $\rho_{0}$
upon further metallization means that scattering on charged
impurities (Fe$^{2+}$ or Fe$^{3+}$ vacancies) is reduced, since they
are screened more effectively with increasing pressure. Note also
the extreme sensitivity of the fitting upon the choice of the
$\rho_{0}$ value. Parenthetically, this sensitivity may be regarded
as an additional criterion of the sample quality.

The well defined exponential behavior of both the pressure and
temperature dependencies of the resistivity match neither those
observed near the quantum critical point for a ferro- and
antiferro-magnetic system,\cite{13} nor those mentioned here for the
antiferromagnetic systems near the critical point for the
Mott-Hubbard transitions.\cite{5,7,9} In other words, the
metallization of Fe$_{3}$O$_{4}$ represents a separate universality
class of phase transitions. In connection, it would be very
interesting to compare them in detail with those\cite{9} for the
spin-canted antiferromagnet NiS$_{2}$, a Mott-Hubbard insulator at
$p=0$, for which the Hund's rule also plays as an important role as
for the ferrimagnetic state of Fe$_{3}$O$_{4}$.

A basic question of relating the observed metallization to the
microscopic properties of the system remains. On phenomenological
grounds, one can relate the thermodynamics of the transition to the
charge-order transition in the octahedral sites, connected with the
freezing of the resonating electron between Fe$^{3+}$-Fe$^{2+}$
configurations of the two ions at $T_{V}$.\cite{14} Considering the
extra electron (Fe$^{2+}$=Fe$^{3+}+e^{-}$) as undergoing Mott-Wigner
localization,\cite{15} one can write down a phenomenological
Ginzburg-Landau functional, albeit with a fermionic form of the
configurational entropy for the spinless (i.e. strongly polarized)
fermions, and rationalize a changeover from the first-to
second-order upon varying the system stoichiometry.\cite{3}

In recent years the above picture was investigated using
spectroscopic methods.\cite{16} A pressure induced transformation
from the inverse to the normal spinel structure was claimed to
occur.\cite{17} This seems to be difficult to reconcile with the
fact, that a transformation of this type should also take place at
$p=0$, which is not observed. In other words, the point $T_{V}(p=0)$
should become also a terminal point of the first-order line
$T_{V}(p)$. Nevertheless, in spite of the complications of the
situation, in which both strong electronic correlations and
electron-lattice coupling are important, a partial charge ordering
was detected by X-ray scattering.\cite{18} This last result means,
that strong correlations may play a primary role in the transition.
This is because a relatively simple exponential behavior of
$T_{V}(p)$, $\rho(T,p_{c})$, and $\rho(T,p>p_{c})$) means, that the
underlying mechanism of the transition should be relatively simple
physically. Otherwise, the change of $T_{V}$ from above $120$~K to
zero should be sizably influenced by e.g. the difference in thermal
occupation of the low-energy phonons and hence lead to nonuniversal
behavior.

\begin{figure}
\includegraphics[width=0.4\textwidth]{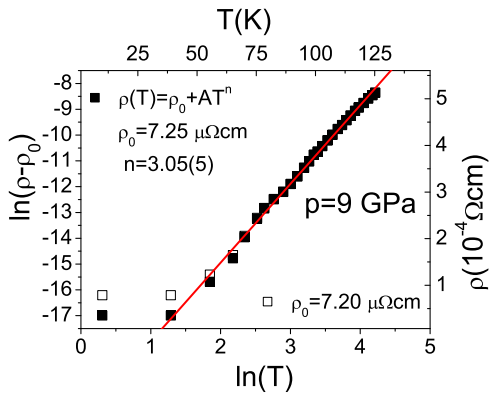}
\caption{\label{fig:f5} Temperature dependence of resistivity
$\rho(T)=\rho_{0}+AT^{n}$ in the
metallic (M) regime for Fe$_{3}$O$_{4.002}$. The
solid line corresponds to $\rho_{0}=7.25\,\mu\Omega$cm and $n=3.05$.
The fitting accuracy does not change under a minute change of $\rho_{0}$,
as shown.}
\end{figure}

Having in mind the underlying simplicity of the electronic
correlations as the primary force, we can easily rationalize the
$T_{V}\sim (p_{c}-p)^{1/3}$ dependence. Namely, regarding strongly
correlated polarized $3d$ electrons as {\em spinless fermions\/}, we
can write down the Landau free energy of the electronic system in
the form\cite{3,14}
$$
F=F_{0}\,+\, E(\Psi)\,-\,k_{B}T[\Psi\ln(\Psi /2)\,+\,
(1-\Psi)\ln (1-\Psi)]
$$
\begin{equation}
\approx F_{0}^{\prime}\,+\, a(T-T_{V})\Psi\, +\, b\Psi^{2}\,+\,\ldots\,,
\end{equation}
where $E(\Psi)=\epsilon_{0}+\epsilon_{1}\Psi
+\epsilon_{2}\Psi^{2}+\dots$ is the system energy, the third term
the relative fermionic entropy, $\Psi$ is the density of itinerant
electrons (resonating Fe$^{2+}\leftrightarrow$ Fe$^{3+}$ electrons
in the octahedral sites), and $\epsilon_{0},\dots,b$ are constants.
The coupling of the electron subsystem to the lattice introduces an
additional term of the Holstein type $\lambda\Psi\Delta x/a=\lambda\Psi (\Delta
V/V)^{1/3}$, where $\Delta x/a$ and $\Delta V/V$ are respectively,
the relative lattice-parameter and volume changes under pressure,
and $\lambda$ is the local electron-lattice coupling constant. 
Assuming that, $\epsilon_{1}\sim\lambda(\ldots )$ and since\cite{12}
$\Delta V \sim p_{c}-p$, the change of
$T_{V}$ induced by applied pressure is $T_{V}=
\lambda^{\prime}(p_{c}-p)^{1/3}$, which is in agreement with
what we observed.

In summary, we have demonstrated, that the metallization of
magnetite under pressure can be described by a simple exponential
scaling of physical properties, including the transition temperature
$T_{V}$, the resistivity near the critical pressure $\rho(p,T_{V+})$
at the transition, and as a function of temperature,
$\rho(p=p_{c},T)$, at $p_{c}$ and above. These properties provide
direct evidence of critical behavior with a quantum critical point
at $p_{c}\approx 8$~GPa. It is also suggested, that the strong
electronic correlations lead to a partial charge order (and
concomitant with it partial localization) and may be the source of
the observed volume change, as is the case for Mott-Hubbard systems.

The work was supported by the Grants Nos. 1 P03B 001 29 and
1 P03B 01 530 from the Ministry of Science
and Higher Education. The work has been carried out partly within the COST P-16 European Project.
We thank also George Honig for discussion.



\begin{thebibliography}{99}

\bibitem{1}
E. J. W. Verwey, Nature {\bf 144}, 327 (1939); E. J. W. Verwey and P. W.
Haayman, Physica {\bf 8}, 979 (1941); for review see: F. Waltz, J. Phys.: Condens. Matter {\bf 14}, R285 (2002).

\bibitem{2}
H. R. Harrison and R. Aragon, Mat. Res. Bull., {\bf 13}, 1097 (1987);
R. Aragon et al.,
Cryst. Growth {\bf 61}, 221 (1983).

\bibitem{3}
J. P. Sheperd etal.,
Phys. Rev. B {\bf 43}, 8461 (1991).

\bibitem{4}
N. M$\hat{\mbox{o}}$ri et al.,
Physica {\bf 312-313}, 686 (2002); S. Todo et al., J. Appl.
Phys. {\bf 89}, 7347 (2001). Metallic behavior was observed
at ambient pressure only above $\approx 300$~K, see: S. Todo et al.,
J. Phys. Soc. Jpn {\bf 64}, 2118 (1995).

\bibitem{5}
P. Limelette et al.,
Science {\bf 302}, 89 (2003).

\bibitem{6}
J. Spa\l{}ek et al., Phys. Rev. Lett. {\bf 59}, 728 (1987);
Phys. Rev. B {\bf 33}, 4891 (1986); {\em ibid\/}, B {\bf 39}, 4175 (1989).

\bibitem{7}
F. Kagawa et al., Nature {\bf 436}, 534 (2005).

\bibitem{8}
A. \'Slebarski and J. Spa\l{}ek,
Phys. Rev. Lett. {\bf 95}, 046402 (2005); J. Spa\l{}ek et al.,
Phys. Rev. B {\bf 72}, 155112 (2005).

\bibitem{9}
A. Husman et al.,
Science {\bf 274}, 1874 (1996); A. Husman et al.,
Phys. Rev. Lett. {\bf 84}, 2465 (2000); for an overview see:
J. M. Honig and J. Spa\l{}ek, Chem. Mater. {\bf 10}, 2910 (1998).

\bibitem{10}
M. Imada et al., Rev. Mod. Phys. {\bf 70}, 1039 (1998).

\bibitem{11}
M. Kobayashi et al., J. Phys. Soc. (Japan) {\bf 55},
4044 (1986) and references therein.

\bibitem{12}
G. Kh. Rozenberg et al., Phys. Rev. Lett. {\bf 96}, 045705 (2006).

\bibitem{13}
T. Moriya, {\em Spin Fluctuations in Itinerant Electron Magnetism\/},
Springer Verlag, Berlin, 1985.

\bibitem{14}
J. M. Honig and J. Spa\l{}ek, J. Less-Common Metals, {\bf 156}, 423 (1989).

\bibitem{15}
N. F. Mott, Festk\"{o}rperprobleme {\bf 19}, 331 (1979).

\bibitem{16}
P. Nov\'{a}k et al., Phys. Rev. B {\bf 61}, 1256 (2000); J. Garcia et al.,
Phys. Rev. B {\bf 63}, 054110 (2001).

\bibitem{17}
M. P. Pasternak et al., J. Magn. Magn. Mat. {\bf 265}, L107 (2003);
H. Kobayashi et al., Phys. Rev. B {\bf 73}, 104110 (2006).

\bibitem{18}
E. Nazarenko et al., Rev. Lett. {\bf 97}, 056403 (2006);
{\bf 98}, 089902 (E) (2007);
see also: H. Schwenk et al., Eur. Phys. J. B {\bf 13}, 491 (2000);
J. P. Wright et al., Phys. Rev. B {\bf 66}, 214422 (2002).

\end{thebibliography}
\end{document}